\documentclass[aps,prl,preprint,superscriptaddress]{revtex4}
\usepackage{graphicx}

\begin{document}
\title{Signature of Chrial Fermion Instability in Ultraquantum Weyl Semimetal TaAs}
\author{Cheng-Long Zhang}
\affiliation{International Center for Quantum Materials, Peking University, Beijing, 100871, China}

\author{Bingbing Tong}
\affiliation{International Center for Quantum Materials, Peking University, Beijing, 100871, China}

\author{Zhujun Yuan}
\affiliation{International Center for Quantum Materials, Peking University, Beijing, 100871, China}

\author{Ziquan Lin}
\affiliation{National High Magnetic Field Center, Huazhong University of Science and Technology, Wuhan, 430074, China}

\author{Junfeng Wang}
\affiliation{National High Magnetic Field Center, Huazhong University of Science and Technology, Wuhan, 430074, China}

\author{Chuan-Ying Xi}
\affiliation{National High Magnetic Field Laboratory, Chinese Academy of Science, Hefei, 230031, China}

\author{Zhong Wang}
\affiliation{Institute for Advanced Study, Tsinghua University, Beijing, 100084, China}
\affiliation{Collaborative Innovation Center of Quantum Matter, Beijing, 100871, China}

\author{Shuang Jia}
\altaffiliation{Electronic address: gwljiashuang@pku.edu.cn}
\affiliation{International Center for Quantum Materials, Peking University, Beijing, 100871, China}
\affiliation{Collaborative Innovation Center of Quantum Matter, Beijing, 100871, China}

\author{Chi Zhang}
\altaffiliation{Electronic address: gwlzhangchi@pku.edu.cn}
\affiliation{International Center for Quantum Materials, Peking University, Beijing, 100871, China}
\affiliation{Collaborative Innovation Center of Quantum Matter, Beijing, 100871, China}



\begin{abstract}
We report the electrical transport properties for Weyl semimetal TaAs in an intense magnetic field.
Series of anomalies occur in the longitudinal magnetoresistance and Hall signals at ultra-low temperatures when the Weyl electrons are confined into the lowest Landau level.
These strongly temperature-dependent anomalies are ascribed to the electron-hole pairing instability.
Our measurements show that the Weyl semimetal TaAs in the ultraquantum regime provides a good platform for studying electron-electron interaction in topological nontrivial  semimetals.
\end{abstract}

\maketitle

A Weyl semimetal (WSM) is a novel topological material whose low energy excited states host the quasiparticles which obey the Hamiltonian for Weyl fermions \cite{Balents-Weyl, Qi-Weyl}.
A WSM provides a unique opportunity for studying the Weyl fermions which are well-known in particle physics literature but have not been observed as fundamental particles in nature.
Weyl semimetals can be realized by breaking a time-reversal symmetry for Dirac semimetals in a magnetic environment, but the semimetals without an inversion center on crystal lattices are possible to host Weyl nodes as well \cite{Wan-Weyl-2011, Burkov-multiWeyl}.
For the type of inversion symmetry breaking Weyl semimetals, the Weyl nodes are the crossing points of a spin-splitted bulk band in momentum space. Weyl fermions appear as quasiparticles of their low-energy excitations near the nodes.
Recently, the first non-magnetic, non-central symmetric WSM tantalum mono-arsenide (TaAs) has been verified in experiments ~\cite{huang_weyl_2015, Weng-TaAs, Xu-TaAs-Weyl, DingHong-TaAs}.
Angle-resolved photoemission spectroscopy (ARPES) experiments for TaAs observed Fermi arcs in its surface electronic states which connect the bulk Weyl nodes with opposite chiralities~\cite{Xu-TaAs-Weyl, DingHong-TaAs}.
Electrical transport measurements in low magnetic fields detected negative longitudinal magnetoresistance (MR) which is ascribed to the Adler-Bell-Jackiw (ABJ) anomaly between the pairs of Weyl nodes ~\cite{Spivak-ABJ, 2015arXiv150302630Z, Chen-Chiral-TaAs}.
Single crystals of TaAs also exhibit ultra-high carrier mobilities and extremely large, linear transversal MR \cite{2015arXiv150200251Z, Chen-Chiral-TaAs}.
All these experimental and theoretical studies on the isostructural family of TaAs open up a new area for the research of topological materials.

In this paper we focus on the electrical transport properties for TaAs in an intense magnetic field ($B$ or $\mu_{0} H$).
For a semimetal with small electron and hole pockets, a strong magnetic field beyond the quantum limit (QL) confines the electrons into the lowest Landau level (LLL) ~\cite{Halperin_QL, Yakovenko-QL, Behnia-graphite-express}.
The LLL for a WSM possesses two linear dispersive subbands which are separated in the momentum space.
When an electric field is applied in the same direction as the magnetic field in this extremity, the charge in each chiral Landau subband is not conserved and chiral anomaly occurs \cite{NielsenABJ, Gorbar-Chiral, Burkov-Weyl-2012, Qi-Weyl}.
If the electron interaction induced phases are not considered, a positive longitudinal magneto-conductivity will occur due to the charge pumping effect between the Weyl nodes with opposite chiralities \cite{NielsenABJ, Qi-Weyl}.
However, a three-dimensional electron system in its LLL exhibits one-dimensional metal characteristic effects \cite{Halperin_QL, Yakovenko-QL}.
Any small interactions will easily induce instabilities of the electrons in such highly degenerate states.
Different types of instabilities in strong magnetic fields have been observed in topologically trivial semimetals \cite{Behnia-graphite-express, Behnia-Bi-Science, Li-Bi-Science}.
The examples include charge density wave (CDW) -like phase transitions and excitonic phases in graphite \cite{Iye-graphite-PRB1982, Yoshioka-graphite-1981, Behnia-graphite-longitudinal,Singleton-graphite, Goto-graphite-1998}.
Electron interaction induced CDW, spin density wave (SDW), and excitonic phases are predicted for Weyl semimetals in theory as well, but the magnetic catalytic phase transitions were less considered \cite{WangZhong-axion,Aji-Weyl-excitonic, Aji-Weyl-PRB2014, Sau-Weyl-QL, RanYing-iridate, WangZhong-Weyl-helix}.
The experiments for the WSM TaAs family are rarely reported in a magnetic field stronger than that in superconducting magnets \cite{shekhar_NbP_2015, 2015arXiv150706301Z}.
All the referred complexity motivates us to study the electrical transport properties of the WSM TaAs beyond its QL.

Band structure calculations and ARPES experiments confirmed that TaAs has two types of Weyl nodes, four pairs of Weyl nodes-1 (W1) on $k_z$ = 2$\pi$/$c$ ($c$ is the lattice constant) plane and the other eight pairs of Weyl nodes-2 (W2) that are away from this plane \cite{huang_weyl_2015, Weng-TaAs, Xu-TaAs-Weyl, DingHong-TaAs}.
In relation to the other three members (NbAs, NbP and TaP) in the isostructural family \cite{Luo-NbAs-PRB, TaPZhanglowfield}, the single crystals of TaAs grown in our lab in general have smaller Weyl electron pockets with the minimal cross section area less than 10 Tesla (T).
The anisotropic Weyl electron pockets close to W1 have the minimal cross section area along the crystallographic $c$ direction, which is three times smaller than that along the $a$ direction.
According to the band structure calculations and ARPES experiments, there also exists trivial pockets which have $2\sim 5$ times larger extremal cross section area than that of the pockets near W1 and W2 \cite{huang_weyl_2015, Xu-TaAs-Weyl}.
The proper chemical potentials for the samples of TaAs provide small enough Fermi pockets which are convenient for exclusively exploring the Weyl electrons' behaviors beyond the QL.

\begin{figure}
 \includegraphics[width=0.8\linewidth]{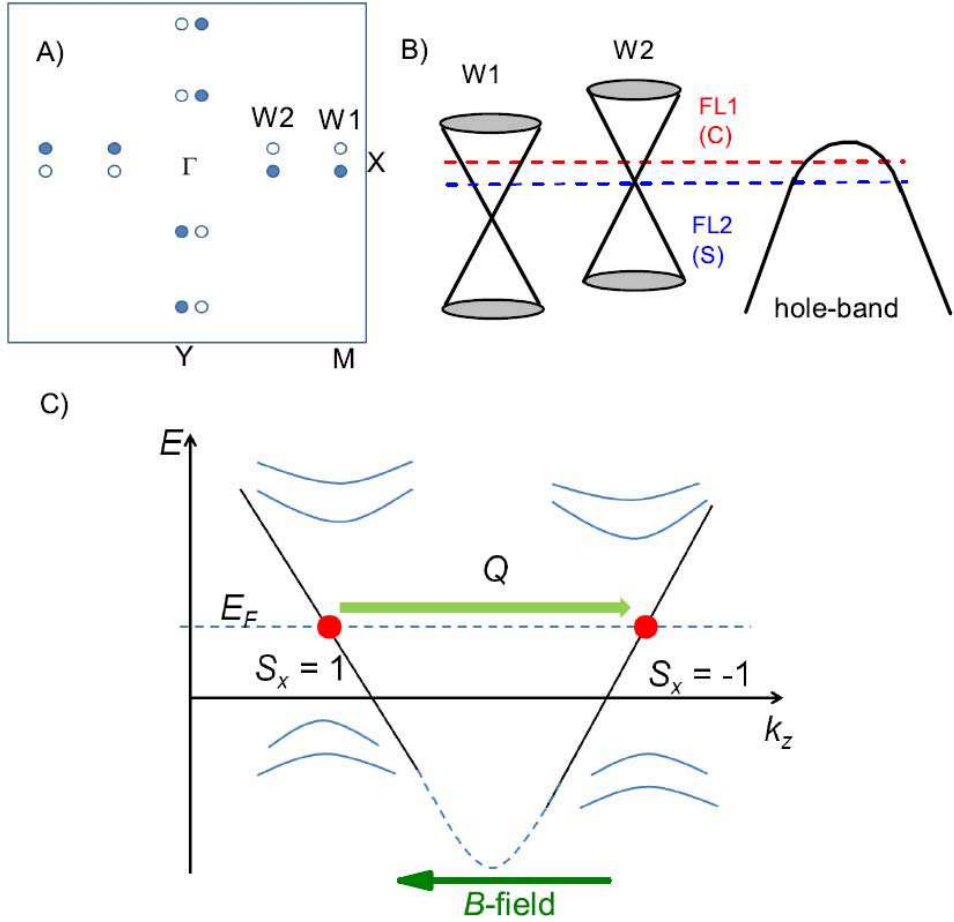}
 \caption{(Color online) (A): a schematic for the projection of all the Weyl points on the $(001)$ surface of Brillouin zone for TaAs. (B): Fermi levels for Sample-S and C. (C): Sketch of the LLL of Weyl fermions and possible vector nesting between the Weyl points with opposite chirality.}
  \label{FIG1}
 \end{figure}

In order to comparably study the electrical properties in ultraquantum region with different directions of the electric field, we chose two single-crystalline samples of TaAs with different configurations: Sample-S has the current along the $a$-axis and the magnetic field along the $c$-axis (Configuration-A: $H\parallel c $, $i\parallel a$), while Sample-C has the electric current and magnetic field parallel to the $c$-axis (Configuration-B: $H\parallel i\parallel c$).
Both of the samples have the Fermi levels close to their Weyl nodes.
Sample-C was mainly measured in a He-3 fridge with a base temperature of 0.35 K in a static magnetic field as high as 35 T in the National High Magnetic Field Lab of China (NHMFLC) at Hefei.
Besides the measurements in a static magnetic field, Sample-S was also measured in an environment with a base temperature of 1.5 K in a pulsed magnetic field as high as 55 T in NHMFLC at Wuhan.

The Shubnikov-de Haas (SdH) oscillations with respect to the reciprocal of the magnetic field for both samples are shown in Panel (B) in Fig. 1 and 2.
The last few Landau levels are labeled as the peaks of the longitudinal resistance after subtracting the background ($\Delta R_{xx}$) and the longitudinal resistivity ($\rho_{zz}$) for Sample-S and C respectively.
We observed a single frequency of $6.8$ T for the SdH oscillations in $\Delta R_{xx}$ and the Hall signals after subtracting the background ($\Delta R_{yx}$) for Sample-S.
For Sample-C there are two sets of oscillations in $\rho_{zz}$ with the frequencies equaling 9.5 and 7.5 T respectively.
The peaks labeled as $N_2$ come from another pocket instead of a set of peaks splitted from $N_1$ due to Zeeman energy because they follow a fixed period of $1/\mu_{0}H$.
Based on the band structure calculations we estimate that the chemical potential is located on 13 meV above W1 for Sample-S where the electron pockets are 8-fold degenerated.
For Sample-C, the chemical potential is located on 18 meV above W1 and the two frequencies are from the extremal cross areas of the electron pockets near W1 and W2.
The magnetic fields stronger than 10 T confine the electrons in the pockets near W1 and W2 to their LLL for both Sample-S and C.

\begin{figure}
 \includegraphics[width=0.8\linewidth]{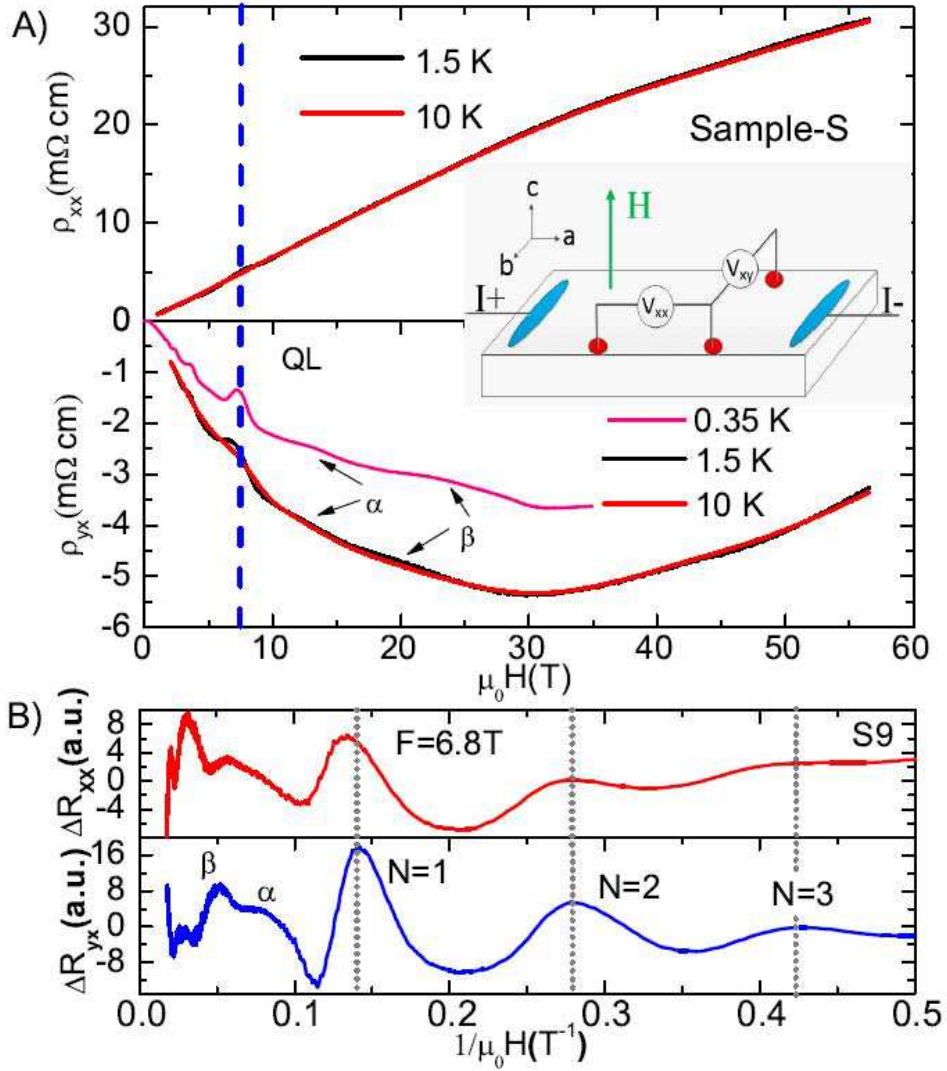}
 \caption{(Color online) Transport measurement results for Sample-S in Configuration-A. (A): $\rho_{xx}$ and $\rho_{yx}$ with respect to the field at $0.35$ K (measured in a static field), $1.5$ and $10$ K (measured in a pulsed field). Different data at $0.35$ K and higher temperatures are due to the different measurement setups in static and pulsed fields. The inset shows Configuration-A. (B): $\Delta R_{xx}$ and $\Delta R_{yx}$, as the data at $1.5$ K subtracted by the data at $10$ K, with respect to the inverse field.}
 \label{FIG2}
 \end{figure}

The field-dependent transversal resistivity ($\rho_{xx}$) and Hall resistivity ($\rho_{yx}$) at $0.35$, $1.5$ and $10$ K for Configuration-A are exhibited in Fig. 2.
The values of $\rho_{xx}$ are proportional to $H$ below 30 T, while the curve of $\rho_{xx}(H)$ bends down very slightly above this field.
The change of $\rho_{xx}$ from 10 K to $1.5$ K is invisible beyond the QL.
On the other hand, the slope of $\rho_{yx}$ changes from negative to positive in a fixed field of 30 T at different temperatures (see more details in SI).
We believe that this slope change is due to the hole contribution coming from a trivial pocket.
Two small bump-like features occur at 14 T and 25 T (labeled as $\alpha$ and $\beta$, respectively) in $\rho_{yx}$ below $1.5$ K but not at 10 K.
The changes of $\rho_{yx}$ for $\alpha$ and $\beta$ are less than 2\%.

\begin{figure}
 \includegraphics[width=0.8\linewidth]{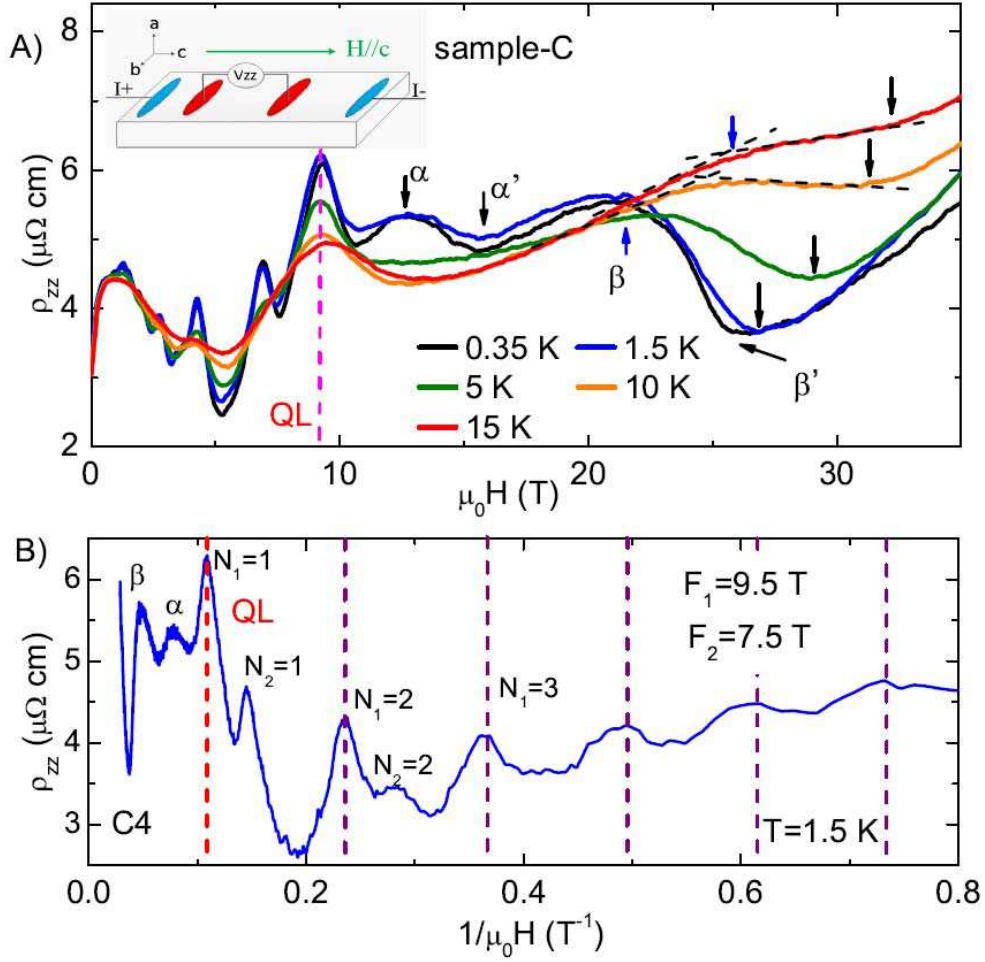}
 \caption{(Color online). Longitudinal resistance for Sample-C in Configuration-B. (A): $\rho_{zz}$ with respect to the field at different temperatures. The inset shows Configuration-B. (B): $\rho_{zz}$ with respect to the inverse field at $1.5$ K.}
 \label{FIG3}
\end{figure}

The longitudinal resistivity in high magnetic fields show much richer features than $\rho_{yx}$ and $\rho_{xx}$.
Disregarding the SdH oscillations which fade out at 15 K, negative field dependent $\rho_{zz}$ of the Sample-C from 1 T to 5 T persists at high temperatures.
This part of negative contribution has been ascribed as an ABJ anomaly of the Weyl electrons \cite{Spivak-ABJ}.
The maximum of $\rho_{zz}$ at the QL of 9 T is also robust at high temperatures when the SdH oscillations have faded out.
Negative longitudinal MR beyond the QL was predicted for Weyl electrons due to a charge pumping effect, which has been observed in Na$_3$Bi and Ag$_2$Se \cite{XiongNa3BiScience, 2015arXiv150706301Z}.

Beyond the QL, $\rho _{zz}$ has two local maxima at 13 T and 22 T at $0.35$ K, which are labeled as  $\alpha$ and $\beta$ respectively.
Two local minima at 15 T and 26 T labeled as $\alpha'$ and $\beta'$ follow $\alpha$ and $\beta$, respectively.
The anomalies of $\alpha$ and $\beta$ occur in the same fields as those on $\rho_{yx}$ in Configuration-A.
The anomalies of $\alpha$ and $\alpha'$ are invisible at 5 K while those of $\beta$ and $\beta'$ persist at 15 K.
Both of $\beta$ and $\beta'$ shift to higher fields with increasing temperature.
Above 15 K, $\rho_{zz}$ monotonically increases with field except the negative part of the contribution near the QL.
The complicated, strongly temperature-dependent behaviors of $\rho _{zz}$ above 10 T are obviously not SdH oscillations, while the impurity scattering beyond the QL does not likely induce such rich features either \cite{Argyres-QL, Halperin_QL}.

We plot $\beta$ and $\beta'$ for Sample-C in the diagram of the magnetic field and temperature (Fig. 4, Panel (A)).
Unlike the fixed field for the QL at 9 T, both of the fields for $\beta$ and $\beta'$ shift to high values with increasing temperature.
The inset of Fig. 4 shows the values of $\rho_{zz}$ at 12.7 T and 26 T at different temperatures.
At 12.7 T $\rho_{zz}$ decreases with increasing temperature below 5 K while $\rho_{zz}$ shows metallic temperature-dependent behavior above 26 T.
It is noteworthy that the change of $\rho_{zz}$ from 0.35 K to 15 K in the ultraquantum region is much more significant than the change of the SdH oscillations below the QL.

A question then arises that what is the mechanism underneath these anomalies above the QL for TaAs.
The different temperature-dependent $\rho_{zz}$ at 12.7 T and 26 T and the changes of the fields for $\beta$ and $\beta'$ at different temperatures suggest that a single particle picture is not able to explain all the features, and interaction-induced instability may be present here.
The instability of the electrons in the ultraquantum region has been investigated in graphite and bismuth \cite{Behnia-Bi}.
Two magnetic field-catalyed phase transitions due to the CDW-like, excitonic instability and depopulation of the Landau subband were observed beyond the QL of graphite \cite{Behnia-graphite-longitudinal,Singleton-graphite}.
These phase transitions induce moderate changes for the transversal MR and four orders of magnitude increase for the longitudinal MR \cite{Behnia-graphite-longitudinal}.
On the other hand, the transversal MR for bismuth shows a weak anomaly far beyond the QL but the origin is still unclear with considering other different probe-adopted experiments \cite{XiongNa3BiScience,Behnia-Bi}.
The transport properties for TaAs are different from those for graphite and bismuth in their ultraquantum region.
Although the transversal MR and the Hall signals for TaAs show very small anomalies, the longitudinal MR shows rich, strongly temperature-dependent features below 15 K beyond the QL, which indicates that an electron-hole pairing instability is likely underneath.

\begin{figure}
 \includegraphics[width=0.8\linewidth]{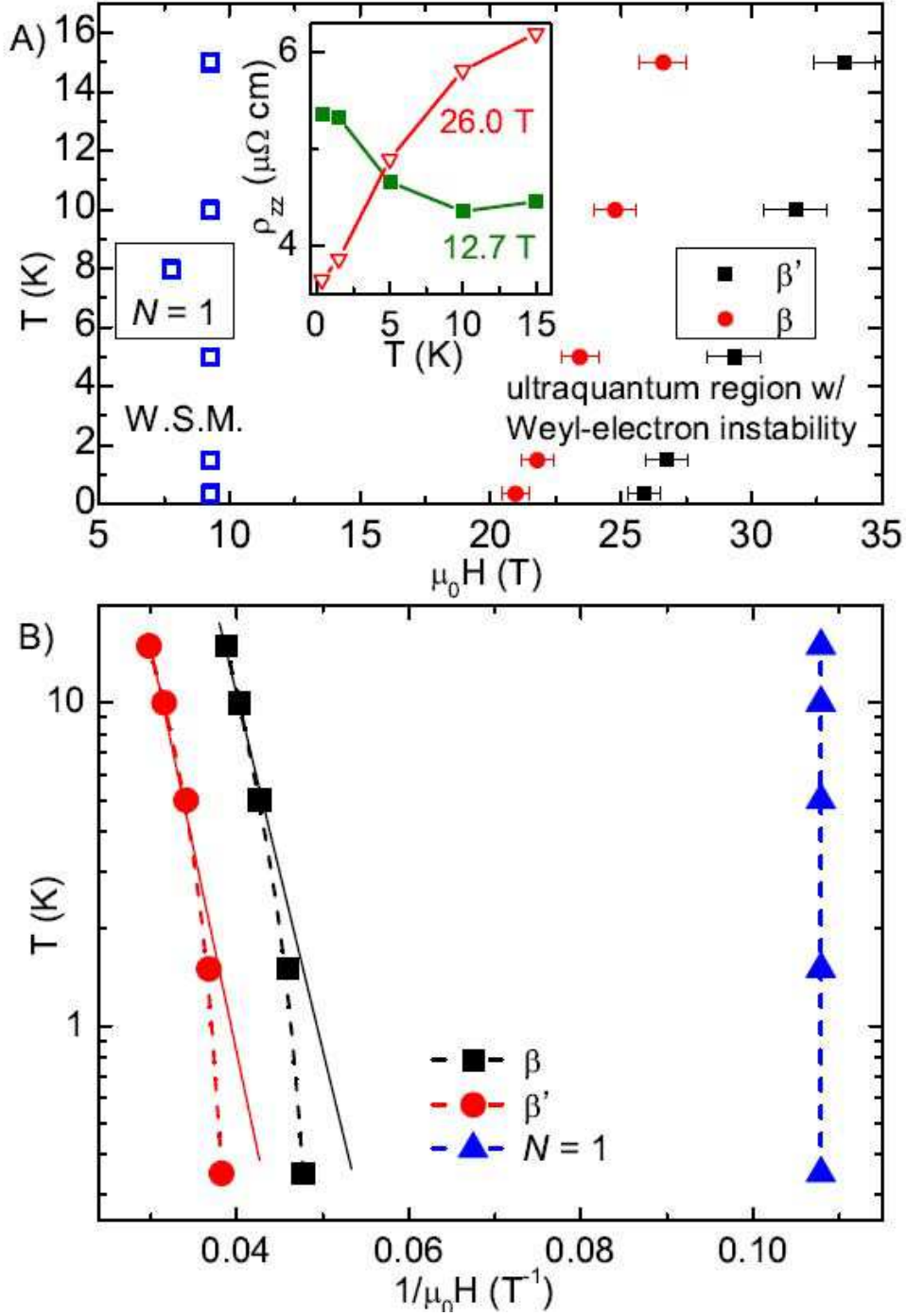}
 \caption{(Color online). (A): The anomalies of $\beta$ and $\beta'$ in $\rho _{zz}$ and the QL  for Sample-C are shown in the $\mu_{0}H-T$ plane. The inset shows the values of $\rho _{zz}$ in $12.7$ T and $26.0$ T at different temperatures. (B): $\beta$ and $\beta'$ and the QL are shown in the $1/\mu_{0}H-T$ plane.}
  \label{FIG4}
 \end{figure}

If we assume an electron-hole-pairing induced phase transition occurs in TaAs, the phase transition temperature $T_{C}$ can be estimated as:
 \begin{equation}
 \label{CDW}
    T_{C} = \Lambda \times exp[-(\frac{\hbar}{eB}) \frac{\hbar v_{F}}{ga^{2}c}]=  \Lambda \times exp(-\frac{B^\star }{B}).
  \end{equation}
in a mean-field approximation of BCS-type instability \cite{Yakovenko-QL, RanYing-iridate}. In this expression, $\Lambda$ is a cutoff energy, which is typically the band width of the Landau band; $a$ and $c$ are the lattice parameters and $g$ is the electron-electron interaction strength, which is estimated to be $1 \sim 2$ eV.
The Fermi wave vector along $c$-axis $k_{F}=\sqrt{\frac{A_{F}}{\pi}}$ and the Fermi velocity $v_{F}$ are estimated as $0.017$~{\AA}$^{-1}$ and $1.31 \times 10^{5}$ m/s respectively by the analysis of the SdH oscillations.
The characteristic magnetic field $B^{*}=(\frac{\hbar}{e}) \frac{\hbar v_{F}}{ga^{2}c}$ from Eq.(\ref{CDW}) is estimated as $\sim 195 - 390 $ T.
This estimation is consistent with $B^{*} \sim 220$ T from the linear fit on the experimental data when $\Lambda \sim 10^4$ K.

In this expression, $T_{C}$ exponentially decays when $B\ll B^{*} $.
As shown in Fig. 4 Panel (B), both temperatures for $\beta$ and $\beta'$ deviate from a linear relationship between $ln T_{\beta}$, $ln T_{\beta'}$ and the inverse magnetic field.
The linear relationship of $ln(T_{C})$ vs. $1/\mu_{0}H$ was reported in pristine graphite, while for neutron-irradiated graphite the relation deviates from linearity which is similar as that for $\beta$ or $\beta'$ \cite{Iye-graphite-impurity, Tokunaga-graphite-excitonic}.
In neutron-irradiated graphite, the deviation is due to the pairing breaking effect by the charged impurities of the lattice defects which significantly affect its $T_{C}$.
TaAs has much more complicated band structure including two different types of pockets near the Weyl points and one trivial hole pocket.
The effect of the trivial hole pocket on the change of $ \rho _{zz}$ is not clear, but we anticipate that $ \rho _{zz}$ will not show a strongly insulating behavior due to the existence of trivial hole conducting channel.
Future experiments such as Nernst signal and magnetic torque measurements in a strong magnetic field will help to elaborate the instabilities for $\alpha $ and $\beta$.

Finally we show that the most plausible instability happening in TaAs is the SDW transitions with the nesting vector {\bf Q} between the two Weyl pockets with opposite chirality.
If the screened Coulomb interaction is responsible for the instability, the interaction should be stronger at smaller nesting vector.
According to the previous band structure calculations and ARPES measurements, the nesting within each pair of Weyl points has the shortest distance from each other in the momentum space.
As shown in Fig. 1 (A), for a pair of W1 near the Brillouin zone boundary located at $(k_{0x}, \pm k_{0y}, 0)$, the nesting vector {\bf Q} equals $(0, 2k_{0y}, 0)$ where $k_{0y} = 0.014\pi/a $ if W1 is initially located at the Fermi level ~\cite{Weng-TaAs}.
In our samples whose Fermi level is away from the Weyl points, the Fermi points of the zeroth Landau level are shifted to $(k_{0x},\pm
k_{0y},\pm k_{0z})$, where $k_{0z}$ can be estimated from the SdH oscillation as $k_{0z}\approx 0.2\pi/c$, assuming that the chemical potential is fixed.
The nesting vector is given by ${\bf Q}=(0,2k_{0y},2k_{0z})\approx (0, 0.03\pi/a, 0.2\pi/c)$.
Due to the opposite spin quantum numbers at the two Fermi points, the instability will generates (see SI) a helical spin density wave ~\cite{WangZhong-Weyl-helix}.
The cases of other three pairs of W1 Weyl points are similar.
The same analysis is also applicable to the W2 Weyl points, whose instability generates helical spin density waves of different forms
(see SI).
From our transport data at hand, it is difficult to distinguish the W1 instability from the W2 instability, and other experimental
tools such as spin-resolved STM may be useful in further investigations.

In summary, we observed anomalies in the electrical transport properties for TaAs in its ultraquantum region.
These strongly temperature-dependent anomalies suggest the signs of the instability of Weyl electrons.
The helical SDWs between the pairs of the Weyl electron pockets with opposite chiralities are the most plausible instability.

\begin{acknowledgments}
We would like to thank Fa Wang, Haiwen Liu and Su-Yang Xu for valuable discussions.
This project is supported by National Basic Research Program of China (Grant Nos. 2013CB921903 and 2014CB921904) and the National Science Foundation of China (Grant No.11374020).
\end{acknowledgments}

\bibliographystyle{unsrt}
\bibliography{TaAshighfield}

\end{document}